\documentstyle[aps]{revtex}

\def\tr{\mathrm{Tr}}
\def\str{\mathrm{Str}}
\draft
\begin{document}

\begin{titlepage}

\begin{flushright}
hep-th/9907036
\end{flushright}

\begin{center}
\vskip3em
{\large\bf Some comments about Schwarzschild black holes
in Matrix theory}

\vskip3em
{J. Kluso\v{n} \footnote{E-mail:klu@physics.muni.cz}}\\ \vskip .5em
{\it Department of Theoretical Physics and Astrophysics\\
Faculty of Science, Masaryk University\\
Kotl\'{a}\v{r}sk\'{a} 2, 611 37, Brno\\
Czech Republic}

\vskip2em
\end{center}

\vfill
\begin{abstract}
In the present paper we calculate the statistical partition function
for any number of extended objects in  Matrix theory in the one loop
approximation. As an to application, we calculate the statistical
properties of K clusters of D0 branes and then the statistical properties
of K membranes which are wound on a torus .
\end{abstract}

\vfill
\end{titlepage}

\section{Introduction}
Matrix theory \cite{Banks1} is one of the most promising ideas for
a nonperturbative definition of M theory. There are many hints
that confirms the validity of this conjecture, for example, this theory can
reproduce the leading order scattering amplitude for gravitons in eleven
dimensional low energy supergravity without momentum transfer in the
longitudinal direction (for example \cite{Becker,Okawa,Taylor4}). 
We also have a good knowledge of what happens
when one compactifies this theory on a $p$ dimensional torus for
$p<6$. \cite{Suskind,Obers,Sen,Seiberg1,Seiberg2} 
This theory also reproduces all five string theories
 \cite{Motl,Seiberg3,Dijkgraaf,Motl2,Krogh} 

On the other hand, there are many unresolved problems. Eleven
dimensional Lorenz invariance is still mysterious, the theory is defined
in a fixed background. Processes with longitudinal momentum transfer
between gravitons has not been calculated and this process would be
rather useful for proving Lorenz invariance. The other serious problem
is that matrix theory is not well understood for more complicated
backgrounds (for example the $p$-torus with $p>6$).

One of the most interesting problems where matrix theory can be
tested is the quantum mechanical description of black holes. This is because
matrix theory, as a candidate for a nonperturbative definition of M
theory, must be able to describe these black holes and indeed this
have been done in many papers (for example \cite{Banks2},
\cite{Martinec}). In these approaches, a remarkably simple model for
Schwarzschild black holes was presented in that the Schwarzschild black
hole is described as a collection of $N$ D0 branes, and the basic
properties of these black holes comes from a statistical study of this
system. In this approach it is important to have these branes
distinguishable in order to obtain the correct formula for the entropy. In
order to obtain distinguishable D0 branes, we must specify some
background, which completely breaks the residual symmetry of the
matrix model, which is the group of permutation of $N$ objects
\(S_N \). More precisely, when we take the Matrix theory Lagrangian,
which describes a system of N D0 branes, this is 0+1 SYM theory with
gauge group U(N). If we want to describe the sector of the theory with
$N$ well separated D0 branes, then we must break the gauge group into
\(U(1)^N \), but the configuration is still invariant under the group of
permutations \( S_N \) which results in the quantum statistic
properties of gravitons (more in \cite{Banks3}), But when we specify some
classical background, this completely breaks the residual symmetry and we
obtain a system of $N$ distinguishable particles \cite{Banks2}.

In \cite{Banks2}, Schwarzschild black holes were studied in
the regime \(S\sim N\), which simply means that the black hole
consists of N distinguishable particles which each carry a
longitudinal momentum \(P_-=1/R \), where R is the radius of the
longitudinal direction in the DLCQ quantization of M theory.

An interesting proposal was made in \cite{Ohta}, where the
statistical properties of Schwarzschild black holes were obtained
{}from first principles in Matrix theory. Because, as the authors
argued, since matrix theory is a candidate for a nonperturbative
description of M theory, it has to be possible to determine from first
principles the properties of all gravitational objects included in
this regime. In their paper, the thermodynamic properties were
calculated from the statistical partition function of Matrix theory
for a particular sector of this theory. The background was chosen to
describe a collection of N classical D0 branes and then they
calculated the partition function by integrating out off diagonal
modes. Subsequently they estimated the basic thermodynamic properties
of this system by the same procedure as was used in \cite{Banks2} for
estimating mean values of characteristic parameters. After this was
done, they obtained the same relation between entropy, mass and
Schwarzschild radius for black hole as in classical gravity and as in
the work \cite{Banks2} without the initial assumption that \( S \sim N
\). In the conclusions of \cite{Ohta}, they suggested a similar way to
count the properties of Schwarzschild black hole also in the regime \(
S<<N \) and they furthermore suggested that in this regime the D0
branes can form membrane-like degrees of freedom.

The regime \( S<<N \) was previously investigated in the paper
\cite{Martinec}
and there it was proposed that the D0 branes would condense into K
clusters, where \(K \sim S \). In this situation these clusters must
also exchange longitudinal momentum among themselves and this
interaction must also be included into the effective potential between
these clusters (as usual in matrix theory, it is difficult to
calculate this potential directly, so the potential has to be
estimated). Doing this they were able to obtain the correct value for
the entropy and the mass of a Schwarzschild black hole in the regime
\( S << N \).

In the present paper we will  calculate the macroscopic properties of
Schwarzschild black holes precisely in this regime \( S<<N \) in a way 
similar to the one suggested in \cite{Ohta}, so this paper should be
seen as a continuation of \cite{Ohta}. The generalization consists
in taking more complicated backgrounds with more than two objects
(there are K extended objects), which are far away from each other. 
We take a background consisting of K clusters of D0 branes
where the longitudinal momentum of the $i$-th cluster is \(P_-=N_i/R
\). We will compute the statistical sum, and after estimating the mean
values of velocity of each object, the distance between the objects
and other parameters according to \cite{Banks2}, we obtain a value for
the entropy \(S \sim K \). This had to be assumed in \cite{Martinec}, but
here it emerges directly from theory. However, we cannot
obtain directly from this approach the correct macroscopic quantities
(mass, entropy, Schwarzschild radius) for black holes due to the fact
that when we count the one loop statistical partition function, we obtain
an effective potential which does not include exchange of longitudinal
momentum between clusters. When we estimate this potential in the
same way as in \cite{Martinec} we obtain correct value of mass,
entropy and Schwarzschild radius for a Schwarzschild black
hole. Certainly it would be interesting to obtain this potential
directly from matrix theory, but we do not know how this could be done.

We also have to worry about the fact that membranes in Matrix theory
are not stable objects unless they are infinite or wound around some
non-trivial cycles. The infinite membranes are not good to construct
finite size black holes from but we will show that the stable compact
membranes wound on non-trivial cycles, which are described by a $3+1$
dimensional SYS theory, can be used to show that the same statements
hold for Schwarzschild black holes in eight dimensions.


This paper is organized as follows. In the first part we choose a
general background and expand the action for matrix theory around it giving
us the action for the fluctuating fields. In the second part 
we evaluate the partition function to one loop order. In the third part
we use our general partition function for a particular
example. Namely, we consider K clusters of D0 branes and we will try
to evaluate the characteristic macroscopic properties of a collection
of these clusters in order to confirm the conjecture \cite{Martinec} that
Schwarzschild black hole in the regime \( S<<N \) can be described as
a collection of K clusters. In the fourth part we calculate this
partition function for matrix theory compactified on a three torus and
as a particular example we take a collection of K membranes wound
around two compact directions and again we obtain the correct formula
for entropy and mass of a Schwarzschild black hole.

\section{General background in matrix theory}
We start with the standard action in {\it DLCQ M theory}, which is
{\it 0+1 SYM} with gauge group $U(N)$   
\begin{eqnarray}
 S=\int dt   \frac{\kappa}{2}\left [ \tr D_t X^n D_t X^n   +   
  \frac{\kappa}{4}g^2\tr\left[X^i,X^j\right]\left[X^i,X^j\right]
  \right.
  & & \nonumber \\ 
           +\left.\frac{\kappa}{2} \tr \left( i \theta^T D_t
  \theta+g \theta^T \gamma_n\left[X^n,\theta\right] \right)\right ] &
  & \nonumber \\ 
\end{eqnarray} 
\begin{equation}
D_tX^n=\partial_t  X^n-i g [ \stackrel{\sim}{A},X^m ] \, , 
 D_t \theta= \partial_t \theta -i g [ \stackrel{\sim}{A}, \theta ]
 \end{equation}
where \( \kappa=\frac{1}{R}\)\, ,\, \(g=M^3R\)  with {\it M} being the
Plank mass and {\it R} being the radius of the compact light-like
circle. Also, \( \gamma^n\) are {\it SO(9)} gamma matrices
in the Majorana representation, so they are real, symmetric and obey
the relation $ \{ \gamma^m , \gamma^n \}=2
\delta_{mn}$, and \(\theta\) is thus a real 16 component spinor. Our
fields also transform in the adjoint representation 
of the gauge group {\it U(N)} so all matrices must be hermitian. 

In order to evaluate the partition function, we must go to Euclidean
time, so we make the transformation 
\begin{equation}
\tau = i t \,  ,  A= - i\stackrel{\sim}{A} 
\end{equation}
After this transformation we obtain the Euclidean form of the action
\begin{eqnarray}\label{act}
S_E=\int d\tau \left[\frac{\kappa}{2}\tr D_{\tau} X^n D_{\tau} 
  X^n- \frac{\kappa g^2}{4} \tr \left[X^n,X^m \right] 
 \left[ X^n,X^m\right]\right. \nonumber  \\ 
  + \left.\frac{\kappa}{2} \tr 
  \left( \theta^T  D_{\tau}\theta-g \theta^T 
  \gamma^m [X^m,\theta]\right) \right] \nonumber  \\
\end{eqnarray}
where \(D_{\tau}X^n=\partial_{\tau}X^n -i g [A, X^n]\) and similarly
for fermions.

We would like to evaluate the partition function, which is defined as
\begin{equation}
Z(\beta)=\int\left[path\right]\exp (-\int_0^{\beta} L_E)
\end{equation}
This function can be evaluated in two steps. To begin with we specify
some background which is described by a classical configuration in
Matrix theory and we evaluate the effective action about this background.
Then, following \cite{Banks2}, we estimate various mean values appropriate for
a background describing a Schwarzschild black hole in eleven
dimensions and insert them in the action to find the quantities we are
interested in. After the evaluation, we find that this model do not
completely describe a Schwarzschild black hole for $N>>S$ and then we
explain why it is so.

We begin by computing the 1 loop effective action for an arbitrary
background. To this end we divide the field from (\ref{act}) into two parts: 
\begin{equation}\label{5}
X^i=\frac{1}{g}B^i+Y^i
\end{equation}
where $X^i$ is the classical background and $Y^i$ is a quantum
fluctuation. We take background values of fermionic fields and ghosts
fields to be zero. 
 Because our Lagrangian is the Lagrangian of SYM theory,
it is gauge invariant and we need to fix the gauge to make the
calculation.
We will use a gauge fixing of the form
\begin{equation}
-\partial_{\tau} A+i [B^n,Y^n] = 0,
\end{equation}
which means that we will also have ghosts present in the
Lagrangian. After expanding the action around the background we obtain
the following Lagrangian (in the following, partial derivative means
derivative with respect to Euclidean time \(\tau\) ):  
\begin{eqnarray} 
L_{\mathrm{boson}}=\frac{\kappa}{2g^2}\tr\partial B^i \partial B^i + 
 \frac{\kappa g^2}{4}\tr\left[B^n,B^m\right]\left[B^n,B^m\right]+ 
 &  & \nonumber \\
 + \kappa \tr\left(\frac{1}{2} \partial Y^i \partial Y^i-2i  
 \partial B^i \left[A,Y^i \right] \right.
 -\frac{1}{2}\left[A,B^m\right]
 \left[A,B^m\right] - & &   \nonumber \\ 
 - \frac{1}{2}\left[B^n,Y^n\right]\left[B^m,Y^m\right] 
 - \frac{1}{2}\left[B^n,Y^m\right]\left[B^n,Y^m\right] & & \nonumber \\
 -\frac{1}{2}\left[B^n,B^m\right]\left[Y^n,Y^m\right] - 
 \left. - \frac{1}{2}\left[B^n,Y^m\right]\left[Y^n,B^m\right]\right) 
  & & \nonumber \\  
L_{\mathrm{fermi}}= \frac{ \kappa}{2} \tr\left( \theta^T \partial_{\tau} 
 \theta -\theta^T \gamma_m \left[B_n, \theta \right] \right) 
 L_{\mathrm{ghost}}=\kappa \tr\left( -\overline{c} \partial^2 c +\overline{c} 
 \left[ B^m, \left[B^m,c \right] \right] \right)
\end{eqnarray}
where the ghost field \(\overline{c}\, , c\) are
two real and independent fermionic fields in the adjoint representation of
the gauge group {\it U(N)}.  

   Now we take the background matrices in the following form
\begin{equation}
 B^m=\left(\begin{array}{cccc}
         B^m_1 & 0    &...    & 0    \\ 
          0 & B^m_2    &...    & 0    \\ 
          0 & 0    &...    & 0    \\ 
          0 &  0  & ...   & B^m_k   \\ 
\end{array} \right) \end{equation}
All other background fields (including gauge field \(A \) ) are chosen
 to be zero.
 The fluctuating fields have the form
\begin{equation}
  A=\left(\begin{array}{ccccc}
         0 & A_{12} & A_{13} &...& A_{1K} \\
         A^{\dagger}_{12} & 0 & A_{23} &... & A_{2K} \\
         ... & ... & ...&... & ...\\
         A^{\dagger}_{1K} & A^{\dagger}_{2K} &...& A^{\dagger}_{K,K-1} & 0 \\
\end{array} \right)\end{equation}
\begin{equation} 
 Y^m=\left( \begin{array}{ccccc}
        0 & Y^m_{12} & Y^m_{13} & ... & Y_{1K} \\
        Y^{\dagger m}_{12} & 0 & Y^m_{23} &... &  Y^m_{2K} \\
         ...& ... & ... & ... & ...\\
         Y^{\dagger m}_{1K} & Y^{\dagger m}_{2K} &... & 
         Y^{\dagger m}_{K,K-1} & 0 \\
\end{array} \right) \end{equation}
where \(X_m\, ,\, m=1\ldots K \) are matrices of  order \( N_m \times N_m
\) and \(Y_{NM} \, ,\, M=1\ldots K-1\, ,\, N=1\ldots K \) are matrices  of
order \( N_M \times N_N\). The fermionic matrix is a fluctuating matrix of the
 form  

\begin{equation}
 \theta =\left( \begin{array}{ccccc}
    0 & \theta_{12} & \theta_{13} & ... & \theta_{1K} \\
    \theta^{\dag}_{12} & 0 & \theta_{23} & ... & \theta_{1K} \\
         ... & ...& ...& ...& ... \\
    \theta^{\dag}_{1K} & \theta^{\dag}_{2K} & ...& \theta^{\dag}_{K,K-1} & 0 \\
\end{array} \right) 
\end{equation} 
We also have ghost fields in the form:
\begin{equation}
c=\left( \begin{array}{ccccc}
                      0 & c_{12} & c_{13} & \dots & c_{1K} \\
                      c^{\dagger}_{12} & 0 & c_{23} & \dots & c_{2K} \\
                      \dots & \dots & \dots & \dots & \dots \\
                      \dots & \dots & \dots & \dots & \dots \\
      c^{\dagger}_{1K} & c^{\dagger}_{2K} & \dots & c^{\dag }_{K,K-1} & 0  \\
\end{array} \right)
\end{equation} 
and the same matrix for \(\overline{c}\), where  the matrices {\it c} are
replaced with \( \overline{c} \). These matrices with indexes \(M,N \) have the
same number of rows and columns as the matrix \(Y_{MN} \).

This form of the matrices of the fluctuating fields is appropriate
for evaluating the one loop effective action because the diagonal
fluctuating fields decouple in the expansion of the Lagrangian at the
quadratic level.
Inserting these matrices in the Lagrangian and after straightforward
but tedious algebra we obtain the following Lagrangian for the
fluctuating fields:
\begin{equation}
L_{boson}=\kappa \left(\sum_{M<N,N=1}^K 
  \left[\dot{Y}^{\dag a}_{MN}\dot{Y}^a_{MN} 
  +Y^{\dag a}_{MN} \left(M^{ab}_{MN} \right)^2 Y^b_{MN}\right] \right)
\end{equation}
where 
\begin{equation}
 Y_{MN}^{\dag a}=(Y^a_{MNij})^{*}
\end{equation} 
is a matrix of order \(N_M \times\ N_N \) , which means that \(i=1
\ldots N_M, j=1\ldots N_N \) 
and the index {\it a} goes 
{}from \(0 \dots  9\) where its zeroth part comes from the field {\it A} and

\begin{equation}
 \left( M^{ab}_{MN} \right)^2=M^{ab}_{{MN}_0} +M^{ab}_{{MN}_1}
\end{equation}
\begin{equation}
M^{ab}_{{MN}_0}=K^2_{MN} \delta^{ab} ,  M^{ab}_{{MN}_1} =-2 i F^{ab}_{MN}
\end{equation}  
\begin{equation}
 K^i_{MN}=B^i_M \otimes 1_{N_N \times N_N} 
  -1_{N_M \times N_M} \otimes{ B^{i}}^T_{MN}
\end{equation}
\begin{equation}
 F^{i0}_{MN}=-F^{0i}_{MN}=\dot{K}^i_{MN} ,\ , 
  F^{ij}_{MN}=i [ K^i_{MN},K^j_{MN} ]
\end{equation}
and by the product between vectors and matrices we mean it in a two
index formulation: 
\begin{equation}
Y^{\dag} CY=({Y}_{ij})^{*}C_{ij,kl}Y_{kl}
\end{equation} 
For the fermions we obtain the following result:
\begin{equation}
L_{fermi}=\kappa \sum_{M<N,N=1}^K\left[ \theta^{\dag \,T}_{MN} 
\dot{\theta}_{MN}- \theta^{\dag \,T}_{MN} \gamma^n K^n_{MN} 
 \theta_{MN} \right]   
\end{equation}      
where \(\theta^{\dag \,T} \theta=({\theta}^T )^{*}_{ij} 
\theta_{ij} \) , because the operation of transposition is related to
the spinor index.

Finally we obtain the contribution from the ghosts:
\begin{eqnarray}
L_{ghost}=\sum_{M<N,N=1}^K \left[ -\overline{c}^{\dag}_{MN}
\partial^2 c_{MN}+\overline{c}^{\dag}_{MN}K^2_{MN}c_{MN} 
+\right. & & \nonumber \\
 \left. +c^{\dag}_{MN}\partial^2 \overline{c}_{MN} 
-c^{\dag}_{MN}K^2_{MN} \overline{c}_{MN} \right]  & & \nonumber \\ 
\end{eqnarray} 

\section{Evaluation of the effective action}  

In the following we would like to calculate the one loop
effective action. Because this subject is well known in the
literature \cite{Tseytlin,Taylor1,Taylor2}, we only briefly
recapitulate basic facts. Firstly, because our effective action is
a thermal effective action, we must specify boundary condition in the
thermal direction for various fields which are in the action. We have: 
\begin{equation}
\Phi (0)=\pm \Phi(\beta)
\end{equation}
where the upper sign is for the bosons (\(Y^i\)) and the lower one for
the fermions ( \( \theta \,, c, \overline{c}\) ).
We use these formulas for evaluating some integrals:
\begin{eqnarray}
\int [ Y] [ Y^{*} ] \exp( (Y^a)^{*} M^{ab} Y^b) 
& =& \det(M)^{-1} \ (\mathrm{for \ bosons}) \nonumber \\
\int [C][ C^{*}] \exp(C^{*}N C ) 
& = & \det(N) \quad (\mathrm{for \ fermions}) \nonumber \\ 
\end{eqnarray}
so
\begin{equation}
\exp{(-\Gamma_1)}=\prod_{M<N,N=1}^K \left(XYZ \right)_{MN}
\end{equation}
where 
\begin{eqnarray}
X_{MN} & = & {\det(-\partial^2+ M_{MN})}^{-1} \nonumber \\
Y_{MN} & = & {\det(-\partial^2 + K^2_{MN})}^2 \nonumber \\
Z_{MN} & = & {\det( \partial-\gamma^n K^n_{MN})}^1  \nonumber \\
\end{eqnarray} 
In these traces we must sum over matrix indices, Lorenz indices or
spinor indices and we must integrate over \(\tau \)  with
appropriate boundary condition. Also, the ghost determinant has exponent
{\it 2}, because we have two equivalent ghosts fields. 

As was explained in \cite{Periwal}, the leading order effective action
can be obtain in an adiabatic approximation 
\begin{equation}
\det(-\partial^2+\omega(\tau)^2) \sim \exp{( \int_{0}^{\beta} \omega(t) dt)}
\end{equation}
We must convert the determinant for fermions into a determinant which
is second order in time derivatives. This can be done by using the fact
that \(\det(\partial +\gamma^n K^n)=\det( \partial-\gamma^n K^n) \),
which gives us
\begin{equation}
F^2=\det(\partial -\gamma^j K^j)\det( \partial +\gamma^i K^i)
 =\det( \partial^2-\frac{1}{2}[K^i,K^j]\gamma^{ij} )
\end{equation}
where \( \gamma^{ij}=\frac{1}{2} [\gamma^i, \gamma^j ] \).

{}From this adiabatic approximation we obtain
\begin{eqnarray}\label{eaction}
\Gamma_1  = -\ln{( \prod_{M<N,N=1}^K (BGF)_{MN})}=  & & \nonumber \\ 
 =-\sum_{M<N,N=1}^K \left[\ln{({\det(B)}^{-1})}
+\ln{({\det(G)}^2)}+\ln{(\det(F)}\right]  = & & \nonumber \\
= \sum_{M<N,N=1}^K \left[ \int_0^{\beta}
\tr( \omega^B(t)_{MN}-\frac{1}{2}\omega^F(t)_{MN}-2 \omega^G(t)) 
\right]  &  &  \nonumber \\ 
\end{eqnarray}
with
\begin{eqnarray}
  {\omega^B(t)}^2_{MN}=M^2(t)_{MN} \ , \ {\omega^F(t)}^2_{MN}
=K^2_{MN}\otimes 1_{16 \times 16}-\frac{i}{2}
F^{ij}_{MN}\gamma^{ij}  \nonumber \\
 \ {\omega^G(t)}^2_{MN}=K^2(t)_{MN} & & \nonumber \\
\end{eqnarray}
In this trace we must sum over spinor or Lorenzian indices and the trace
 in the matrices  \(M_{MNij}\) is over the \( i , j \) indices. 
Now we must evaluate this effective action. This will be done following
\cite{Taylor1}. We must divide, the various \(\Omega(t)\) into two
parts, the first part which scales as a function of distance between
two long separate object and the rest. This can be done in the
following way; for bosons: 
\begin{eqnarray}
{\Omega(t)^{ab}}^2_{MN}={M_0}^{ab}_{MN} +{ M_1}^{ab}_{MN} \nonumber \\
M(t)^{ab}_{0MN}=K(t)^2_{MN} \delta^{ab} 
& , &  M(t)^{ab}_{1MN}=-2 i F^{ab}_{MN} \nonumber \\
\end{eqnarray}
for fermions:
\begin{eqnarray}
\Omega(t)^2_{MN}=M(t)_{0MN}+M(t)_{1MN} & &  \nonumber \\
M(t)_{0MN}=K(t)^2_{MN} \otimes 1_{16 \times 16}  
& , &  M(t)_{1MN}= \partial K(t)^n_{MN} \gamma^n 
- \frac{1}{2} F^{ij}_{MN} \gamma^{ij}  \nonumber \\
\end{eqnarray} 
We must go over to the proper time representation :
\begin{equation}
\tr \sqrt{M(t)_{0MN} +M(t)_{1MN}}=- \frac{1}{2\sqrt{ \pi}} 
\tr \int_{0}^{\infty} \frac{d \tau}{{ \tau}^{ \frac{3}{2}}}
\exp(-\tau \left( M(t)_0 +M(t)_1\right)) 
\end{equation} 
The integral can now be evaluated by the help of a Dyson perturbative
series. Because we know, that \(M_1\) is much smaller than \(M_0\), we
can take the latter as a perturbative term, and the former as a
background around which we expand. It corresponds to the standard
Dyson formula free Hamiltonian. We take (in the following we will
not write out the dependence of these terms on t, because in the proper
time representation these terms are constant. We also will not write
various indices, because these are not important in what follows) 
\begin{eqnarray}
U(s)=\exp^{-s(M_0+N_1)} ,  \  U(s)_0= \exp^{ -sM_0} & & \nonumber  \\
V(s)=U(-s)_0U(s) \ , \ \frac{dV(s)}{ds}=-M(s)_1V(s) \, \ 
M(s)_1=\exp^{i s M_0} M_1 \exp^{-i s M_0} \nonumber \\
\end{eqnarray} 
\begin{eqnarray}
V(s)=1- \int_{0}^s M(s_1)_1ds_1+\int_0^s M(s_2) 
\int_0^{s_2}M(s_1)_1ds_1ds_2 \dots \\
U(s)=U(s)_0 V(s) , \ \tr\sqrt{M_0+M_1}
=-\frac{1}{2 \sqrt{\pi}}\tr \int_{0}^{\infty} 
\frac{ds}{s^{\frac{3}{2}}} U(s)  \nonumber \\
\end{eqnarray}

Zero order:
\begin{equation}
\tr_L(1_{10 \times 10})-\frac{1}{2}\tr_L(1_{16 \times 16} )-2=0
\end{equation}
where in evaluating we divide the trace into two parts, one over the
Lorenz indices and the other over the matrix indices. In the following
we will not write indices of type \(MN\), because as we can see from
(\ref{eaction}),  
the effective action is linear in indexes \(M,N \) so that we can
simply calculate  
for one matrix with index \(MN\) and then sum over \(M,N \).
 We see that to zeroth order these contributions are
zero.
To first order, we have contributions from boson and fermion respectively:
\begin {equation}
\tr_LM(s)_{1B}-\tr_LM(s)_{1F}= 
  -2 i F^{aa}+\partial K^n \tr \gamma ^n 
  -\frac{i}{2} F^{ij} \otimes \tr \gamma^{ij} =0
\end{equation}
This is due to the basic properties of the gamma matrices \ ( \(\tr
\gamma^n=0\) ) and the antisymmetry of tensor \(F^{ab}\). 

To second order, we obtain the following term in the effective action:
\begin{equation}
-\frac{1}{2 \sqrt{2}}\tr_G \int_0 ^{\infty}  
  \frac{ds}{s^{ \frac{3}{2}}} \exp^{-sM_0}
  \tr_L\left( \int_0^s ds_1M(s_1) \int_0^{s_1} ds_2 M(s_2)\right)
\end{equation}
The bosonic term has the form:
\begin{equation}
\tr_L(M(s_1)_1M(s_2)_1)=8\dot{K}(s_1) \dot{K} (s_2) -4 F^{ij}(s_1)F^{ij}(s_2)
\end{equation}
and the fermionic term has the form:
\begin{eqnarray}
\tr_L(M(s_1)_1M(s_2)_2)=  \nonumber \\
  = \left( \dot{K}^n(s_1) \gamma^n -\frac{i}{2} F^{ij}(s_1) 
  \gamma^{ij} \right) \left( \dot{K}^m(s_2)\gamma^m
  -\frac{i}{2}F^{kl}(s_2) \gamma^{kl} \right) =   \nonumber \\
 =  \dot{K}^n(s_1) \dot{K}^m(s_2)\delta^{mn} -\frac{1}{4} 
  F^{ij}(s_1) F^{kl} (s_2) \tr(\gamma^{ij} \gamma^{kl} )  =   \nonumber \\
 =   16 \dot{K}(s_1) \dot{K} (s_2) -8 F^{ij}(s_1)F^{ji}(s_2)   \nonumber \\
\end{eqnarray}
so fermionic and bosonic contribution cancel each other.
To third order we obtain:
\begin{equation}
\frac{1}{\sqrt{2} \sqrt{\pi}}
  \int_0^{\infty}\frac{ds}{s^{\frac{3}{2}}}
  \exp^{-sM_0}\int_0^s ds_1 M(s_1)_1\int_0^{s_1}ds_2 M(s_2)_1 
  \int_0^{s_2}ds_3 M(s_3)ds_3
\end{equation}
which for bosons means:
\begin{eqnarray}
\tr_L\left(M(s_1)_1 M(s_2)_1 M(s_3)_1\right) = \nonumber \\
=8 i \left(-\dot{K}^i(s_1)F_{ij}(s_2)
 \dot{K}^j(s_3)-F^{ij}\dot{K}^j(s_2)\dot{K}^i(s_3)-\right. \nonumber \\
\left. -\dot{K}^j(s_1) \dot{K}^i(s_2)F^{ij}(s_3)+
  F^{ij}(s_1)F^{jk}(s_2)F^{ki}(s_3) \right) \nonumber \\
\end{eqnarray} 
and for fermions:
\begin{eqnarray}
\tr_L\left(M(s_1)_1M(s_2)_1M(s_3)_1\right)
=\tr_L\left(\dot{K}^n \gamma^n- \frac{i}{2}F^{ij} 
 \gamma^{ij} \right) (s_1) \times
 \nonumber \\
\times \left(\dot{K}^m \gamma^m -\frac{i}{2} 
 F^{kl}\gamma^{kl} \right) (s_2) \left( \dot{K}^p 
 \gamma^p-\frac{i}{2}F^{op} \gamma^{op} \right)(s_3)=  \nonumber \\
\end{eqnarray}
where the trace is now over spinor indices. After some algebra we
obtain the following result for the fermions:
\begin{equation}
=\left(-16 i F^{ij}F^{jk}F^{ki}- 16i 
 \dot{K}^i F^{ij}\dot{K}^j -16i F^{ij}\dot{K}^i 
 \dot{K}^j +16 i F^{ij}F^{jk}F^{ki} \right)(s_1)(s_2)(s_3)
\end{equation}
so again, to third order, fermionic and bosonic term cancel each other.
 The first nontrivial term we obtain at fourth order, so
when we take the leading terms of \(M_1(s)\), which is independent of
\(s\), we obtain the following result
\begin{eqnarray}
\tr_{L \ boson}\left( M(s_1)_1M(s_2)_1M(s_3)_1M(s_4)_1\right)_
  {MN} \nonumber \\
-\frac{1}{2} \tr_{L \ fermi}\left(M(s_1)_1 M(s_2)_1 
  M(s_3)_1 M(s_4)_1\right) _{MN}=  \nonumber \\
=F_{MN}=\str(24 F_{oi}F_{oi}F_{oj} F_{oj} 
  -24 F_{0i} F_{oi} F_{jk} F_{jk}  -  \nonumber \\
-96 F_{oi} F_{oj} F_{ik}F_{kj} +24F_{ij} F_{jk}F_{kl} 
  F_{li} -6 F_{ij}F_{ij} F_{kl} F_{kl} ) \nonumber \\
\end{eqnarray} 
where $\str$, meaning the symmetric trace, is defined as an average
over all ordering in indices \(ij\) and to leading order we may
replace \(M_0\) in the exponential function in our Dyson series by the
term \(r_{MN} \) so after a simple integration, and because the trace
is independent to leading order in \(s\) we obtain the final form for
the effective 
potential: 
\begin{equation}
\Gamma_1=\frac{-5}{128}\sum_{M<N , \ N=1}^K \int_0^{\beta}  
  \frac{dt}{r^7_{MN}} \str F(t)_{MN}
\end{equation}
This is the final result, which can be used for studying the thermodynamic
properties of Schwarzschild black holes.

\section{Schwarzschild black holes}

We will follow \cite{Banks2} in supposing that we have some bound state
corresponding to some classical object in matrix theory which move in
any bounded region in space time and then we will try to evaluate the
thermodynamic characteristics of this system. We will take our background
in the form of \(K\) clusters of \(D0 \) branes which are
far away from each other and we will suppose that their  quantum
mechanical properties are not important for obtaining the leading
order result. So we will take the background in the following form: 
\begin{equation}
X^m= \left( \begin{array}{ccc}
(v^m_1 \beta +r^m_1) \otimes 1_{N_1\times N_1} & 0   \\
0 & (v^m_2 \beta +r^m_2) \otimes 1_{N_2 \times N_2} &  0  \\

0 &   \dots & ( v^m_K \beta +r^m_K) \otimes 1_{N_K \times N_K} \\
\end{array} \right)
\end{equation} 
where \(r_i^m \) describes position of the cluster in time \(\beta=0 \)
and where \(v^m_i \) is the velocity of the cluster and \(m=1 \dots 9
\). By choosing a background of this form we do not care about the
proper structure of the bound state (cluster) of these
particles. Inserting these background matrices into the effective
action we obtain: 
\begin{equation}
\Gamma_0=\beta \sum_{M=1}^K \frac{1}{2R} N_M v^2_M
\end{equation}
\begin{equation}
\Gamma_1=-\frac{15}{16} \sum_{M<N, N=1}^K \int_0^{\beta} dt 
 \frac{N_NN_Mv^4_{MN}}{(v^2_{MN}{t}^2+r^2_{MN})^{\frac{7}{2}}}
\end{equation}
Now we must make a scale transformation, because this effective action
has been evaluated for background matrices \(B^n\) and these are
related to the physical coordinates and physical velocities through
the following transformations:
\begin{equation}
B^i=gX^i, \ g=M^3_p R, \ v \rightarrow g v, \ \Gamma_1 
 \rightarrow \frac{1}{g^3} \Gamma_1
\end{equation}
After this transformation we obtain as a final result:
\begin{equation}
\Gamma= \Gamma_0 + \Gamma_1=\beta  \sum_{N=1}^K \frac{N_N v^2_N}{2R} 
-\frac{15 l^9_P}{16 R^3}\sum_{M<N, \ N=1}^K \int_0^{\beta} 
dt \frac{N_MN_M v^4_{MN}}{(v^2_{MN} {t}^2 + r^2_{MN} )^{\frac{7}{2}}}
\end{equation}
Now we would like to evaluate some thermodynamic properties for this
ensemble. We will suppose that these clusters live in some bounded
region in space-time and that the radius of this  region is
\(R_s\). Since these clusters must live in this region, their thermal
average value must obey the following conditions:
\begin{equation}
\left < |r_{MN}| \right > \sim \left< |v_{MN}| \right> 
 \beta \sim \left<|v_M| \right> \beta \sim R_s
\end{equation}
where \( \beta \) is the inverse temperature of the black hole.
The cluster must obey the Heisenberg uncertainty principle, which in
our situation says
\begin{equation}
\frac{\left<N_M \right>}{R} \left< |v_M| \right> R_s \sim 1
\end{equation}
because the momentum of the cluster (as can be derived from the
Lagrangian, using Noether's theorem, since our Lagrangian is invariant
under translation) is
\begin{equation}
 P^i_M= \frac{N_M v_M}{R}
\end{equation}
We can also suppose that all clusters have on the average the same
value of longitudinal momentum \(P_-\), which tells us that
\begin{equation}
\left<N_M \right> \sim \frac{N}{K}
\end {equation}
When we combine all these facts, we obtain following average values:
\begin{equation}
\left<|v_M|\right> \sim \left< |v_{MN} | \right> 
\sim \frac{RK}{N R_s} ,\  \beta \sim \frac{R^2_s N}{RK}
\end{equation} 
Now, since these average values have been estimated, we can start to
count the statistical properties of this ensemble. We have 
\begin{equation}
Z( \beta)=\sum_{over parametrs} exp ^{-(\Gamma_0 +\Gamma_1)}
=\left<exp^{-(\Gamma_0+ \Gamma_1)( \beta)} \right>
\end{equation}
\begin{equation}
\left< E \right>= -\frac{ \partial \ln Z( \beta) }{ \partial \beta }
  =\left< \frac{ \partial ( \Gamma_0 +\Gamma_1 ) }{ \partial \beta } \right>
\end{equation}
when we apply our average values, we obtain
\begin{equation}\label{energy}
\left<E \right> \sim \frac{R K^2 N}{2 R^2_s N^2} 
-\frac{15 G_{11} R K^4}{32 N^2 R^{11}_s}
\end{equation}
The Helmholtz free energy is given by
\begin{eqnarray}
F=-\frac{\ln Z( \beta) }{ \beta} =  \nonumber \\
=\left< \sum_{N=1}^K \frac{N_N v^2_N}{2R} 
-\frac{15 G_{11}}{16 R^3} \sum_{M<N, N=1}^K \frac{1}{\beta} 
\int_0^{\beta}dt
\frac{N_NN_Mv^4_{MN}}{(v^2_{MN}t^2+r^2_{MN})^{\frac{7}{2}}} 
\right> \nonumber \\
\end{eqnarray}
{}From this free energy we can obtain the entropy:
\begin{eqnarray}
S={\beta}^2 \frac{\partial F(\beta) }{ \partial \beta } 
= \left< \frac{15 G_{11}}{16 R^3} \sum_{M<N,N=1}^K \int_0^{\beta} 
\frac{N_NN_M v^4_{NM}}{(v^2_{MN} t^2 +r^2_{MN})^{\frac{7}{2}}} 
- \right. \nonumber \\
- \left. \beta \frac{15 G_{11}}{16 R^3} \sum_{M<N, N=1}^K 
\frac{N_NN_M}{(v^2_{MN} {\beta}^2+r^2_{MN})^{\frac{7}{2}}} 
\right> \nonumber \\
\end{eqnarray} 
Inserting our average values into the previous term, we obtain the
entropy in the following form
\begin{equation}
S \sim \frac{G_{11} K^3 N^2} {R^9_s N^3}
\end{equation}
In order to obtain the value for the radius \(R_s\), we apply the
virial theorem, which has the form 
\begin{eqnarray}\label{radius}
\left<E_{\mathrm{kin}} \right> \sim \left< E_{\mathrm{pot}} \right>
 \rightarrow \nonumber \\
\rightarrow \frac{R K^2 N}{R^2_s N^2} \sim \frac{ G_{11} 
R K^3}{N^2 R^{11}_s } \rightarrow R_s \sim 
\left( \frac{K^2 G_{11}}{N} \right)^{\frac{1}{9}} \nonumber \\
\end{eqnarray}
When we use the previous result for estimating the entropy, we obtain
\begin{equation}
S \sim K << N
\end{equation}
So our result confirms the conjecture of \cite{Martinec}, that
Schwarzschild black holes in region \(S << N \) can be described as a
bound state of \( K \) clusters, but this is not the end of the story.

When we insert (\ref{radius}) into (\ref{energy}) we obtain
\begin{equation}
 E \sim R G^{-\frac{2}{9}}_{11} K^{ \frac{14}{9}} N^{-\frac{7}{9}}
\end{equation}
and when we use the relation between mass and light-cone energy, we
obtain a final result for the mass of a Schwarzschild black hole:
\begin{equation}
M^2=2 P_-E \rightarrow M \sim \left( G^{-1}_{11}
 N \right)^{\frac{1}{9}} K^{\frac{7}{9}}
\end{equation}
which certainly is not the correct result, since mass is a function of
the number \(N \), but mass is a macroscopic quantity which
cannot depend on microscopic parameters. The reason why we cannot
obtain a correct result lies in our ignorance of the process of
longitudinal momentum exchange. But this is difficult problem in
matrix theory, which cannot be resolved with our simple result. But
if we follow \cite{Martinec} and estimate the interaction term in
the way that was used in that paper we obtain
\begin{equation} 
E_{\mathrm{pot}} \sim -\frac{N}{K} \left< \Gamma_1 \right>  
\end{equation}
so the average value of the energy is
\begin{equation}\label{energy1}
\left<E \right> \sim \frac{R K^2N}{2 R^2_s N^2} 
-\frac{15 N G_{11} R K^4}{32 K N^2 R^{11}_s}
\end{equation}
and when we apply virial theorem to (\ref{energy1}), we obtain 
\begin{equation} \label{radius1}
R_s \sim \left( G_{11} K \right) ^{\frac{1}{9}}
\end{equation}
When we count entropy in the same way as before, we obtain
\begin{equation}
S \sim \frac{G_{11} N^3 K^3}{R^9_sK N^3} \sim \frac{G_{11} K^2}{R^9_s} \sim K
\end{equation}
and the energy
\begin{eqnarray}
\left< E \right> \sim \frac{R K^2 N}{2 R^2_s N^2} 
-\frac{15 G_{11} N K^4 R}{32 K N^2 R^{11}_s } \sim \nonumber \\
\sim \frac{R}{N} \left( \frac{K^8}{G_{11}}\right )^{\frac{2}{9}} \nonumber \\
\end{eqnarray}
and the  mass of Schwarzschild black hole becomes
\begin{equation}\label{mass1}
M=\sqrt{ \frac{N}{R} E} \sim \left( \frac{K^8}{G_{11}}\right)^{\frac{2}{9}}
\end{equation}
and finally, when we insert (\ref{mass1}) into (\ref{radius1}), we obtain
\begin{equation}\label{radius2}
R_s \sim \left(G_{11} M \right)^{\frac{1}{8}}
\end{equation}
which is the correct result for the Schwarzschild radius as a function
of the mass of the black hole. Finally, we will determine the temperature of
the Schwarzschild black hole. To begin with we write the transformation
rule in DLCQ for a transformation in the longitudinal direction:
\begin{equation}
[M_{+-},P_+]= i P_+ , \ [M_{+-}, P_-] =-i P_-
\end{equation}
which means, that these two quantities scale with relatively inverse
coefficient under this transformation. And because temperature
transforms as an energy, we have:
\begin{eqnarray}
P_-=k P_{0-} \rightarrow k=\frac{P_-}{P_{0-}} , \ (P_-=\frac{N}{R}, 
  \ P_{0-}=\frac{M}{ \sqrt{2}} ) \nonumber \\
\frac{1}{ \beta} =k^{-1} T_0 \rightarrow T_0=
  \frac{ \sqrt{2} N}{R M \beta} \sim \frac{1}{R_s} \nonumber \\
\end{eqnarray}
again this is a correct result.

\section{Objects in Matrix theory on a torus}

So far we have described only flat eleven dimensional matrix theory
and general configuration matrix objects in this theory. We would
like to confirm the conjecture of \cite{Ohta}, that the
Schwarzschild black hole in the regime \(S<<N \) can be described
as a collection of S membranes. However, there is a
problem. The previous effective action is certainly valid formally,
but cannot describe extended objects in matrix theory for finite
N. This is because these objects are described by commutator  \(
F^{ij}=i [K^i,K^j ] \) terms, of which the trace is certainly
zero for finite matrices. These extended objects should therefore be
described by infinite matrices, but this is not the end of the story.
Even classically, membranes are stable only if they are
infinite, or if they are living on some compact manifold. Finite
membranes in flat space-time will eventually collapse into a point in
the classical case, but when their size will be of the order of the
Planck scale, the quantum mechanical properties will be important so
then we must study the whole system as a quantum mechanical system
which is described by a special \(0+1\) SYM with gauge group \(U(N)\)
and this is a rather difficult problem.
On the other hand, classical infinite
membranes cannot serve as a model for Schwarzschild black hole. 
In order to describe a statistical ensemble of stable extended
objects (for example membranes), we must take these extended objects as
being infinite or we must wind them on some compact dimension. In
order to describe a membrane, which is finite, we must take  Matrix
theory on a compact manifold, 
and the simplest way is to compactify Matrix theory on a p-torus,
which by the standard prescription \cite{Taylor3} 
is the same as SYM on the dual p-torus. In the following we will study
membranes which are wound over a 3-torus so Matrix theory is described
by 3+1 SYM with gauge group \(U(N) \), for a system with total longitudinal
momentum \( P_-=  \frac{N}{R} \). 
The bosonic part of this Lagrangian has the form:
\begin{equation}\label{lag}
L= \frac{1}{g^2} \int d^{p}\tilde{x} \tr( -\frac{1}{4} F_{ \mu \nu}
F^{ \mu \nu}  
+ \frac{(D_{\mu} \phi)^2}{2} + [ { \phi} ^i ,{ \phi}^j ]^2 )
\end{equation} 
where the integration is over the dual 3 torus of size (for simplicity
all compact dimensions have the same size) $\Sigma = \frac{ 2 \pi
l^3_p}{R L}$. Here \(R \) is the radius of the longitudinal direction
and \(L\) is the radius of the compact dimension of the original torus. Greek
indices run over \(0 \dots 3\) and over \( i,j =4 \dots 9\). In
the Yang-Mills theory, the coupling constant is 
\begin{equation} \label{coupling}
 g^2= \frac{l^3_p}{L^3} .
\end{equation}
We would like to show how the action for the membrane arise from the
previous Lagrangian since the generalization for more membranes is
straightforward. (It simply consists in breaking gauge group \(U(N) \)
into \( U(N_1) \times U(N_2) \times \dots U(N_K) \), where each
subgroup describes one particular membrane.) 

We would like to describe a membrane moving in the fourth direction,
and which is wound in the first and second direction. In order to
describe this membrane we take a classical background in the following
way. We suppose that \( \phi \) are position independent and that
there are no Wilson lines \( \oint A_{ \mu} d x^{ \mu} =0 \). Then the
kinetic term for the scalar field reduces to the form \[ \frac{1}{g^2}
\tilde{V} \dot{ \phi} ^2 \] where \( \tilde{V} \) is the volume of the
dual torus which is related to the volume of the original torus V through
the standard form $ \tilde{V} = \frac{l^9_p (2 \pi )^3}{R^3 V }$. We
must also relate this scalar field with the physical momentum 
of the membrane in the fourth direction. This also means that the only
nonzero field is \( \phi _4 \), all the others are zero. We take this
field to be of the form : 
\begin{equation}\label{back}
\dot{ \phi^i}= K^i \otimes 1_{N \times N}
\end{equation}
The unknown constant K can be determined from this condition. The
action with (\ref{lag}) is invariant under the transformation:
\begin{equation}
 \phi ^{'i}= \phi^i  + b^i \otimes 1_{N \times N} 
\end{equation}
where \(b^i\) is an arbitrary constant. This invariance 
 is nothing but Poincare translation symmetry in the
transversal space. By using standard Noether methods we obtain
conserved quantities which are just the moment in the transverse
dimensions:
\begin{equation}\label{impuls}
P^i=  \frac{1}{g^2}\int \tr \dot{ \phi^i} dx= \frac{1}{g^2} 
  \tilde{V} \tr \dot{ \phi^i}
\end{equation}
and finally, when we insert our background ansatz (\ref{back}) into
(\ref{impuls}) and using (\ref{coupling}), we obtain 
\begin{equation}
P=\frac{K N L^3 l^9_p (2 \pi )^3}{l^3_p R^3 L^3} 
\Rightarrow K=P \left( \frac{R}{l^3_p (2 \pi )^3} \right)^2 \frac{1}{P_-}
\end{equation}
As a check, we may insert (\ref{back}) into the Hamiltonian, which has
the form
\begin{equation}\label{hamiltonian}
H=\frac{1}{g^2} \int \tr ( \frac{1}{4} F_{ \mu \nu} F^{ \mu \nu }
 + \frac{1}{2} \dot{ \phi } ^2)
\end{equation}
and if we for the moment leave out the gauge field term, we obtain
\begin{equation}
H=\frac{1}{g^2} \tr \dot{ \phi}^2  \tilde{V} =\frac{P^2 R}{N} = \frac{P^2}{P_-}
\end{equation}
which is the correct result.

In order to describe a membrane, we must also have a non-vanishing
value of magnetic flux and when the membrane is wound on the first and
second direction, this flux is
\begin {equation}
\frac{1}{2 \pi } \int \tr F_{12} dx^{1}dx^{2}=n
\end{equation}
where the integral is over the first and second dimension and \( n\)
is an integer. Again, when our background configuration is independent
of the space coordinates, we obtain the following value for \( F_{12} \) 
\begin{equation}
F_{12}= \frac{n}{N} \frac{ 2 \pi}{ \Sigma ^2} = \frac{n}{N} 
 \frac{ R^2 L^2 }{2 \pi l^6_p}
\end{equation}
and when we insert the previous result into (\ref{hamiltonian}), we obtain 
\begin{equation}
\frac{1}{2g^2} F^2_{12} \tilde {V}= \frac{R}{2N} n^2 ( 2 \pi ) 
 \frac{L^4}{l^6_p}= \frac{ 2 \pi M^2}{2P_-}
\end{equation}
{}from which we obtain the correct mass of a membrane which is \(n \)
times wound around the first and second direction.
\begin{equation}\label{mass}
M= \frac{L^2n}{l^3_p}
\end{equation} 
Now when we take the background in the following form:
\begin{equation}
F_{12}= \left( \begin{array}{cccc}
F^1 \otimes 1_{N_1 \times N_1} & 0 & \dots  & 0 \\
0 & F^2 \otimes 1_{N_2 \times N_2}  & \dots & 0 \\
\dots & \dots & \dots  & \dots \\
0 & \dots & \dots & F^K \otimes 1_{N_K \times N_K } \\
\end{array} \right)
\end{equation}
where we have
\begin{equation}
\frac{1}{2 \pi} \int \tr F^i_{12} = \frac{1} {2 \pi} \tilde{V}NF^i =n^i
\end{equation}
and 
\begin{equation}
\phi ^4= \left( \begin{array}{cccc}
 k^1t \otimes 1_{N_1 \times N_1} & 0 & \dots & 0 \\
0 & k^2t\otimes 1_{N_2 \times N_2 } & \dots & 0 \\
\dots & \dots & \dots & \dots \\
0 & \dots & \dots & k^Kt \otimes 1_{N_K \times N_K }
\end{array} \right)
\end{equation}
and insert it into the effective action, we obtain the effective
action for K membranes in the form:
\begin{equation}
\Gamma_0= \beta \sum_{i=1}^K ( \frac{P^2_i}{2P_{i-}} +\frac{2 \pi
M^2_i}{2 P_{i-}} )
\end{equation} 
We express this effective action in momentum variables because this is
more appropriate for compactification (we can take the momentum to be
finite), but from (\ref{mass}) we see that in the limit \( L
\rightarrow 0 \), mass goes to zero, so that the membrane looks like
it lives in a non-compact dimension as a D0 brane. 

Now we may start to analyze the effective potential between the K
membranes described by the previous background. This
potential was obtained in a previous work \cite{Tseytlin}, which
in our case, $3+1 \; SYM$,  takes the form:
\begin{equation}\label{potencial}
\Gamma_1= - \sum_{M<N,N=1}^K \frac{a}{r^4_{MN}}\int_0^{ \beta}d 
 \tilde{x} \str(F_{MN})
\end{equation}
where the minus sign is due to the fact that we are calculating the finite
temperature effective potential which means that we are working in
Euclidean signature and
\begin{eqnarray}
F_{MN}=\left(24F_{oi}F_{oi}F_{oj}F_{oj} 
 - 24F_{oi}F_{oi}F_{jk}F_{jk}\right. &  & \nonumber \\
-96F_{oi}F_{oj}F_{ik}F_{kj} +24F_{ij}F_{jk}F_{kl}F_{li} & &  \nonumber \\
\left. -6F_{ij}F_{ij}F_{kl}F_{kl} \right)_{MN} & & \nonumber \\
\end{eqnarray}
For our background we obtain these nonzero terms
\begin{eqnarray}
F_{12MN}=F_{12M} \otimes 1_{N \times N} +1_{M \times M} 
    \otimes F^T_{12N} \nonumber \\
F_{04MN}=\dot{ \phi }_{4M} \otimes 1_{N \times N} - 1_{M \times M} 
  \otimes \dot{ \phi}_{4N} \nonumber \\
\end{eqnarray}
When we insert this background into (\ref{potencial}) and using 
\begin{equation}
\tr\left( A_M \otimes A_N B_M \otimes B_N \dots \right) 
  =\tr \left(A_MB_M \dots \right) \tr \left(A_NB_N \dots \right)
\end{equation}
and using the fact that for a commuting background (such as in our case)
$\str$ is the same as an ordinary trace, we obtain
\begin{eqnarray}
\Gamma_1= -\frac{A G_8}{R} \sum_{M<N, N=1}^N \int_0^{ \beta}dt 
  \frac{P_{-M}P_{-N}}{r_{MN}^4} \times \nonumber \\
\times \left[ \left( \frac{P_M}{P_{-M}}- \frac{P_N}{P_{-N}} 
  \right)^2 - \frac{L^2}{(2 \pi)^2} \left( \frac{n_N}{N_N} 
  +\frac{n_M}{N_M} \right)^2 \right] ^2 \nonumber \\
\end{eqnarray}
where \( A\) is a numerical constant, which is not important for our
purposes $G_8=\frac{l^9_p}{L^3}$, and the meaning of the other
symbols was explained in the previous part where we considered one
simple membrane. Again we see that in the limit $L \rightarrow 0$
the last term in the effective potential is zero, so again this
effective action reduces to the effective action of K D0 branes in
eight dimensions:
\begin{eqnarray}
\Gamma= \Gamma_0 + \Gamma_1 = \nonumber \\
 \beta \sum_{i=1}^K \frac{P^2_i}{P_{-i}}-\frac{AG_8}{R} 
  \int_0^{\beta}dt \frac{P_{-M}P_{+N}}{r(t)^4_{MN}}
  \left( \frac{P_M}{P_{-M}}- \frac{P_N}{P_{-N}} \right)^4 \nonumber \\
\end{eqnarray}
We see that again this configuration describe an ensemble of K D0
particles, so that the analysis will be the same as in the previous
case or as in \cite{Martinec}, so that we will not repeat this
analysis there.

\section{Conclusion}
In the previous parts we have done some simple calculations in  
Matrix theory. The basic goals of this paper was to confirm the conjecture
in \cite{Ohta}, that a Schwarzschild black hole in the regime $ N>>S $ can be
described as a collection of K membranes which each has a longitudinal momentum
$ P_-\sim N/KR $. This goal was achieved in the sense that we managed
to derive, from first principles, that the entropy is proportional to
the number of membranes. By estimating the correct potential
energy between these clusters we also managed to derive the correct
values for the other parameters of the black hole.
We also showed that the same results hold when we use membrane
constituents that are stable against collapse, namely, which are wound
around some compact manifold.

Clearly, it would be of great interest to also calculate the potential
between the clusters from first principles. This, however, is a
process that involves longitudinal momentum transfer which is a
notoriously difficult problem in matrix theory so we will leave this
problem for the future.

\section*{Acknowledgements }

 I would like to thank Rikard von Unge and Zde\v{n}ek Kopeck\'{y} for 
helpful discussions.

\end{document}